\documentclass[english,preprint,superscriptaddress]{revtex4-1}
\usepackage[T1]{fontenc}
\usepackage[latin9]{inputenc}
\usepackage{color}
\usepackage{amssymb}
\usepackage{graphicx}
\usepackage{esint}

\makeatletter

\providecommand{\tabularnewline}{\\}

 \@ifundefined{textcolor}{}
 {%
   \definecolor{BLACK}{gray}{0}
   \definecolor{WHITE}{gray}{1}
   \definecolor{RED}{rgb}{1,0,0}
   \definecolor{GREEN}{rgb}{0,1,0}
   \definecolor{BLUE}{rgb}{0,0,1}
   \definecolor{CYAN}{cmyk}{1,0,0,0}
   \definecolor{MAGENTA}{cmyk}{0,1,0,0}
   \definecolor{YELLOW}{cmyk}{0,0,1,0}
 }

\makeatother

\usepackage{babel}
\begin{document}

\title{Magnetic reconnection mediated by hyper-resistive plasmoid instability }

\author{Yi-Min Huang}

\email{yiminh@princeton.edu}

\selectlanguage{english}%

\affiliation{Center for Integrated Computation and Analysis of Reconnection and
Turbulence }

\affiliation{Center for Magnetic Self-Organization in Laboratory and Astrophysical
Plasmas}

\affiliation{Max Planck-Princeton Center for Plasma Physics and Princeton Plasma
Physics Laboratory, Princeton, NJ 08543}

\author{A. Bhattacharjee}

\affiliation{Center for Integrated Computation and Analysis of Reconnection and
Turbulence }

\affiliation{Center for Magnetic Self-Organization in Laboratory and Astrophysical
Plasmas}

\affiliation{Max Planck-Princeton Center for Plasma Physics and Princeton Plasma
Physics Laboratory, Princeton, NJ 08543}

\author{Terry G. Forbes}

\affiliation{Space Science Center, University of New Hampshire, Durham, NH 03824}
\begin{abstract}
Magnetic reconnection mediated by the hyper-resistive plasmoid instability
is studied with both linear analysis and nonlinear simulations. The
linear growth rate is found to scale as $S_{H}^{1/6}$ with respect
to the hyper-resistive Lundquist number $S_{H}\equiv L^{3}V_{A}/\eta_{H}$,
where $L$ is the system size, $V_{A}$ is the Alfv\'en velocity,
and $\eta_{H}$ is the hyper-resistivity. In the nonlinear regime,
reconnection rate becomes nearly independent of $S_{H}$, the number
of plasmoids scales as $S_{H}^{1/2}$, and the secondary current sheet
length and width both scale as $S_{H}^{-1/2}$. These scalings are
consistent with a heuristic argument assuming secondary current sheets
are close to marginal stability. The distribution of plasmoids as
a function of the enclosed flux $\psi$ is found to obey a $\psi^{-1}$
power law over an extended range, followed by a rapid fall off for
large plasmoids. These results are compared with those from resistive
magnetohydrodynamic studies.
\end{abstract}
\maketitle

\section{Introduction}

Magnetic reconnection is arguably one of the most important processes
in plasma physics, which provides a mechanism to release the energy
stored in magnetic field and convert it to thermal energy or bulk
plasma kinetic energy. It is generally believed to be the underlying
mechanism that powers explosive events such as solar flares, magnetospheric
substorms, and sawtooth crashes in fusion plasmas.\citep{ZweibelY2009,YamadaKJ2010}
The key challenge of magnetic reconnection theory applied to these
events is how energy can be released explosively in time scales that
are very short compared with the characteristic resistive diffusion
time scale.

Traditionally, it was widely accepted that magnetic reconnection in
the resistive magnetohydrodynamics (MHD) model is described by the
classical Sweet-Parker theory.\citep{Sweet1958,Parker1957} Sweet-Parker
theory predicts that reconnection rate scales as $S^{-1/2}$, where
$S\equiv V_{A}L/\eta$ is the Lundquist number (here $V_{A}$ is the
upstream Alfv\'en speed, $L$ is the reconnection layer length, and
$\eta$ is the resistivity). Because the Lundquist number $S$ is
usual very high (e.g. in solar corona $S\sim10^{12}-10^{14}$, assuming the classical Spitzer resistivity), the
Sweet-Parker reconnection rate is too slow to account for energy release
events. For this reason, research on fast reconnection in the past
two decades has mostly focused on collisionless reconnection, which
can yield reconnection rates as fast as $\sim0.1V_{A}B$ (here $B$
is the upstream magnetic field).\citep{BirnDSRDHKMBOP2001} In order
to trigger collisionless reconnection, the current sheet width has
to first go down to kinetic scales such as the ion skin depth or ion
thermal gyroradius.\citep{Aydemir1992,MaB1996a,DorelliB2003,Bhattacharjee2004,CassakSD2005,CassakDS2007} 

The super-Alfv\'enic plasmoid instability,\citep{LoureiroSC2007}
which is a secondary tearing instability acting on a Sweet-Parker
current sheet, has drawn considerable interest in recent years. While
the scaling features of this linear instability are surprising in
their own right,\citep{LoureiroSC2007,BhattacharjeeHYR2009,SamtaneyLUSC2009,NiGHSYB2010}
what is more important is that the instability leads to a nonlinear
regime in which the reconnection rate becomes nearly independent of
$S$,\citep{BhattacharjeeHYR2009,HuangB2010,UzdenskyLS2010,LoureiroSSU2012}
in striking departure from the prediction of Sweet-Parker theory.
Furthermore, the plasmoid instability causes fragmentation of the
Sweet-Parker current sheet to smaller secondary current sheets, which
allows collisionless reconnection to be triggered earlier than previously
thought possible.\citep{DaughtonRAKYB2009,ShepherdC2010,HuangBS2011} 

Over the past few years, the linear analysis of the plasmoid instability
mediated by resistivity has been extended by various authors to include
Hall,\citep{BaalrudBHG2011} three dimensional (3D),\citep{BaalrudBH2012}
and shear flow effects.\citep{LoureiroSU2013} Nonlinear evolution
of the instability has also been extensively studied in two dimensional
(2D) systems.\citep{BhattacharjeeHYR2009,CassakS2009,HuangB2010,BartaBKS2011,ShenLM2011,LoureiroSSU2012,NiZHLM2012,NiRLZ2012,MeiSWLMR2012,Baty2012}
The resistive plasmoid instability in 2D is now relatively well understood.
However, the resistive tearing mode is not the only mechanism that
causes plasmoid formation. In fact, plasmoid formation has been found
to be ubiquitous in large scale reconnection simulations, regardless
of the underlying physical models. \citep{DrakeSSRK2006,DaughtonSK2006,DaughtonRAKYB2009,DaughtonRKGYAB2010,DaughtonRKYABB2011,FermoDS2012}

In this work, we explore the consequences of hyper-resistivity on
magnetic reconnection mediated by the plasmoid instability. In this
model, the Ohm's law assumes the form $\mathbf{E}=-\mathbf{u}\times\mathbf{B}-\eta_{H}\nabla^{2}\mathbf{J}$,
where $\eta_{H}$ is the hyper-resistivity. The origin of hyper-resistivity
has been attributed to anomalous electron viscosity due to micro-scale
field line stochasticity as well as tearing-mode turbulence.\citep{FurthRS1973,KawVR1979,BhattacharjeeH1986,Strauss1988a,BhattacharjeeY1995,CheDS2011,Biskamp1993}
This paper is organized as follows. The linear theory of the instability
is established in Sec. \ref{sec:Linear-Theory}, where the scaling
of the linear growth rate is derived. Sec. \ref{sec:Numerical-Simulations}
presents the results from nonlinear simulations. First we verify the
scaling of the linear growth rate, then move on to study the effects
on magnetic reconnection when the instability has evolved into fully
nonlinear regime. We focus on scalings of reconnection rate, the number
of plasmoids, and the sizes of secondary current sheets. We give a
heuristic justification for the scalings. Finally, we examine the
statistical distribution of the magnetic flux contained in plasmoids,
which is a topic of considerable interest in recent years. The results
from both linear and nonlinear studies are summarized and discussed
in Sec. \ref{sec:Summary-and-Discussion}, and comparisons are made
with the plasmoid instability in resistive MHD.

\section{Linear Theory\label{sec:Linear-Theory}}

The linear analysis of the plasmoid instability in resistive MHD was
first carried out by Loureiro\emph{ et al.}\citep{LoureiroSC2007}
The analysis shows that the maximum growth rate $\gamma_{max}$ scales
as $S^{1/4}V_{A}/L$, and the number of plasmoids scales as $S^{3/8}$.
Subsequently, it was shown that these scalings emerge directly from
the classic tearing mode dispersion relation,\citep{CoppiGPRR1976}
by taking into account the property that the width of the Sweet-Parker
current sheet $\delta_{CS}$ scales as $L/S^{1/2}$.\citep{BhattacharjeeHYR2009,HuangB2013}
Here we follow the latter approach for the hyper-resistive plasmoid
instability. To do that we need two ingredients: a generalization
of the Sweet-Parker theory and a linear tearing mode theory with hyper-resistivity
in place of resistivity. 

The generalization of Sweet-Parker theory\citep{Sweet1958,Parker1957}
with hyper-resistivity is straightforward. Let $u_{i}$ and $u_{o}$
be the inflow speed and the outflow speed, respectively. The conditions
for conservation of mass and energy remains unchanged, which give
\begin{equation}
u_{i}L\sim u_{o}\delta_{CS},\label{eq:mass-conservation}
\end{equation}
\begin{equation}
\rho u_{o}^{2}\sim B^{2}.\label{eq:energy-conservation}
\end{equation}
Here $\rho$ is the plasma density, $B$ is the magnetic field, and
we have neglected numerical factors of $O(1)$. The only deviation
from the resistive Sweet-Parker theory comes from the Ohm's law, which
is now $\mathbf{E}=-\mathbf{u}\times\mathbf{B}-\eta_{H}\nabla^{2}\mathbf{J}$.
Under quasi-steady condition, the out-of-plane electric field is spatially
uniform, which gives the following condition
\begin{equation}
u_{i}B\sim\eta_{H}\nabla^{2}J\sim\eta_{H}\frac{B}{\delta_{CS}^{3}},\label{eq:induction}
\end{equation}
where we have made use of the relations $J\sim B/\delta_{CS}$ and
$\nabla^{2}\sim1/\delta_{CS}^{2}$. From Eqs. (\ref{eq:mass-conservation})
-- (\ref{eq:induction}), the following scaling relations are obtained:
\begin{equation}
u_{o}\sim\frac{B}{\sqrt{\rho}}\sim V_{A},\label{eq:outflow}
\end{equation}
\begin{equation}
\delta_{CS}\sim\frac{L}{S_{H}^{1/4}},\label{eq:width}
\end{equation}
and
\begin{equation}
u_{i}\sim S_{H}^{-1/4}V_{A},\label{eq:inflow}
\end{equation}
where $S_{H}$ is he hyper-resistive Lundquist number defined as 
\begin{equation}
S_{H}\equiv\frac{L^{3}V_{A}}{\eta_{H}}.\label{eq:SH}
\end{equation}

The linear dispersion relation of hyper-resistive tearing mode has
been derived by Aydemir.\citep{Aydemir1990} Here we outline the analysis
to keep this paper self-contained. For simplicity, the plasma is assumed
to be incompressible with a uniform density. In two-dimensional (2D)
Cartesian coordinates $(x,z)$, the plasma flow $\mathbf{u}$ and
the magnetic field \textbf{$\mathbf{B}$} can be expressed in terms
of the stream function $\phi$ and the flux function $\psi$ as $\mathbf{u}=\nabla\phi\times\mathbf{\hat{y}}$
and $\mathbf{B}=\nabla\psi\times\mathbf{\hat{y}}$, and the system
can be described by the well-known reduced MHD equations.\citep{KadomtsevP1974,Strauss1976,Biskamp1993}
Consider an equilibrium magnetic field $\mathbf{B}=B_{x}(z)\mathbf{\hat{x}}$,
with $V_{A}(z)$ being the corresponding Alfv\'en speed profile.
Assuming linear perturbations of the form $\tilde{\phi}=\tilde{\phi}(z)e^{ikx+\gamma t}$
and $\tilde{\psi}=\tilde{\psi}(z)e^{ikx+\gamma t}$, where $\gamma$
is the growth rate and $k$ is the wavenumber along the $x$ direction,
the linearized reduced MHD equations with hyper-resistivity are: 
\begin{equation}
\gamma\mathcal{D}\tilde{\phi}=ikV_{A}\mathcal{D}\tilde{\psi}-ikV_{A}''\tilde{\psi},\label{eq:phi}
\end{equation}
\begin{equation}
\gamma\tilde{\psi}=ikV_{A}\tilde{\phi}-\eta_{H}\mathcal{D}^{2}\tilde{\psi}.\label{eq:psi}
\end{equation}
Here primes denote $d/dz$ and the operator $\mathcal{D}\equiv d^{2}/dz^{2}-k^{2}$.
We further assume that $z=0$ is the resonant surface, where $V_{A}=0$.

Away from $z=0$, the effect of hyper-resistivity is negligible. Therefore,
in the outer region $\tilde{\phi}\simeq\gamma\tilde{\psi}/ikV_{A}$,
which can be used in Eq. (\ref{eq:phi}) to eliminate $\tilde{\phi}$.
It follows that if we assume $\gamma^{2}\ll k^{2}V_{A}^{2}$, the
plasma inertia (left hand side of Eq. (\ref{eq:phi})) is negligible.
Hence, in the outer region, the perturbed flux function $\tilde{\psi}_{o}$
is governed by 
\begin{equation}
\left(\mathcal{D}-\frac{V_{A}''}{V_{A}}\right)\tilde{\psi}_{o}\simeq0.\label{eq:outer}
\end{equation}
The solution of Eq. (\ref{eq:outer}) is subject to appropriate outer
boundary condition, e.g. $\tilde{\psi}_{o}\to0$ at infinity if the
global length scale of the domain along the $z$ direction under consideration
is much larger than the current sheet width. In general, the solutions
of $\tilde{\psi}_{o}$ from both regions $z>0$ and $z<0$ will not
match smoothly at $z=0$. The mismatch in the slope of $\tilde{\psi}_{o}$
is characterized by the tearing stability index \citep{FurthKR1963}
\begin{equation}
\Delta'\equiv\left.\frac{\tilde{\psi}_{o}'}{\tilde{\psi}_{o}}\right|_{0^{-}}^{0^{+}},\label{eq:deltaprime}
\end{equation}
which is completely determined by the equilibrium profile and the
wave number $k$. 

Because the outer solutions do not match smoothly at $z=0$, a boundary
layer exists around $z=0$. In the inner region around $z=0$, a separate
set of boundary layer equations 
\begin{equation}
\gamma\tilde{\phi}''=ikV_{A}'(0)x\tilde{\psi}'',\label{eq:inner1}
\end{equation}
\begin{equation}
\gamma\tilde{\psi}=ikV_{A}'(0)x\tilde{\phi}-\eta_{H}\tilde{\psi}''''\label{eq:inner2}
\end{equation}
are solved. Asymptotic matching of the inner and outer solutions give
the linear growth rate. Interested readers are referred to Ref. \citep{Aydemir1990}
for a detailed asymptotic analysis. It is instructive, however, to
employ a simple heuristic argument that gives correct scalings of
the linear growth rate, as follows.

Let $a$ be the width of the equilibrium current sheet and $\delta$
be the width of the boundary layer. Tearing modes are often classified
into the so-called constant-$\psi$ and nonconstant-$\psi$ regimes,
depending on whether $\tilde{\psi}$ is approximately constant or
not within the boundary layer. In the constant-$\psi$ regime, the
variation of $\tilde{\psi}'$ is approximately $\tilde{\psi}\Delta'$
across the boundary layer. Hence, we may estimate $\tilde{\psi}''\sim\tilde{\psi}\Delta'/\delta$
and $\tilde{\psi}''''\sim\tilde{\psi}\Delta'/\delta^{3}$. Also $\tilde{\phi}''$
may be estimated as $\tilde{\phi}/\delta^{2}$, and $V_{A}'(0)\sim V_{A}/a$.
Balancing terms in Eqs. (\ref{eq:inner1}) and (\ref{eq:inner2})
yields

\begin{equation}
\gamma\tilde{\phi}/\delta^{2}\sim k\frac{V_{A}}{a}\delta\frac{\Delta'\tilde{\psi}}{\delta},\label{eq:const-psi1}
\end{equation}
\begin{equation}
\gamma\tilde{\psi}\sim k\frac{V_{A}}{a}\delta\tilde{\phi}\sim\eta_{H}\frac{\Delta'\tilde{\psi}}{\delta^{3}}.\label{eq:const-psi2}
\end{equation}
Solving Eqs. (\ref{eq:const-psi1}) and (\ref{eq:const-psi2}) yields
the scalings of $\delta$ and $\gamma$. The results are 
\begin{equation}
\delta\sim S_{Ha}^{-2/9}(\Delta'a)^{1/9}(ka)^{-2/9}a\label{eq:const-psi-delta}
\end{equation}
and
\begin{equation}
\gamma\sim S_{Ha}^{-1/3}(\Delta'a)^{2/3}(ka)^{2/3}(V_{A}/a),\label{eq:const-psi-gamma}
\end{equation}
where 
\begin{equation}
S_{Ha}\equiv\frac{a^{3}V_{A}}{\eta_{H}}\label{eq:SHa}
\end{equation}
is the hyper-resistive Lundquist based on the length scale of the
equilibrium current sheet width $a$. For the commonly employed Harris
sheet profile with $V_{A}\propto\tanh(z/a)$, the tearing stability
index is given by 
\begin{equation}
\Delta'=\frac{2}{ka^{2}}(1-(ka)^{2}).\label{eq:deltaprime-Harris}
\end{equation}
Using Eq. (\ref{eq:deltaprime-Harris}) in Eqs. (\ref{eq:nonconst-psi-delta})
and (\ref{eq:nonconst-psi-gamma}) yields
\begin{equation}
\delta\sim S_{Ha}^{-2/9}(ka)^{-1/3}(1-(ka)^{2})^{1/9}a\label{eq:delta-Harris}
\end{equation}
and 
\begin{equation}
\gamma\sim S_{Ha}^{-1/3}(1-(ka)^{2})^{2/3}(V_{A}/a)\label{eq:gamma-Harris}
\end{equation}
in the constant-$\psi$ regime. More precisely, the $O(1)$ numerical
factor in Eq. (\ref{eq:gamma-Harris}) can be determined by an asymptotic
matching calculation,\citep{Aydemir1990} which gives
\begin{equation}
\gamma\simeq(2/\pi)^{2/3}S_{Ha}^{-1/3}(1-(ka)^{2})^{2/3}(V_{A}/a).\label{eq:gamma-harris1}
\end{equation}

In the nonconstant-$\psi$ regime, $\tilde{\psi}$ varies significantly
within the boundary layer, and we may estimate $\tilde{\psi}''\sim\tilde{\psi}/\delta^{2}$
and $\tilde{\psi}''''\sim\tilde{\psi}/\delta^{4}$. Following the
same procedure of balancing terms in Eqs. (\ref{eq:inner1}) and (\ref{eq:inner2})
yields

\begin{equation}
\delta\sim S_{Ha}^{-1/5}(ka)^{-1/5}a\label{eq:nonconst-psi-delta}
\end{equation}
and 
\begin{equation}
\gamma\sim S_{Ha}^{-1/5}(ka)^{4/5}(V_{A}/a).\label{eq:nonconst-psi-gamma}
\end{equation}

The transition wavenumber from the constant-$\psi$ regime to the
nonconstant-$\psi$ regime may be estimated from the self-consistency
of the constant-$\psi$ assumption, as follows. The variation of $\tilde{\psi}$
within the boundary layer may be estimated as $\Delta\tilde{\psi}\sim\tilde{\psi}'\delta\sim\tilde{\psi}\Delta'\delta$.
Therefore, $\tilde{\psi}$ being approximately constant requires $\Delta\tilde{\psi}\ll\tilde{\psi}$,
i.e. $\Delta'\delta\ll1$. From Eqs. (\ref{eq:deltaprime-Harris})
and (\ref{eq:nonconst-psi-delta}), the self-consistency criterion
$\Delta'\delta\ll1$ requires $ka\gg S_{Ha}^{-1/6},$ assuming $S_{Ha}^{-1/6}\ll1$.
Therefore, we expect that the transition occurs when $ka\sim S_{Ha}^{-1/6}$.
Note that $\gamma$ is a monotonically decreasing function of $ka$
in the constant-$\psi$ regime, and a monotonically increasing function
of $ka$ in the nonconstant-$\psi$ regime. At the transition wavenumber
the growth rates from two branches coincide, which gives the peak
growth rate $\gamma_{max}\sim S_{Ha}^{-1/3}(V_{A}/a).$

\begin{figure}
\begin{centering}
\includegraphics[scale=0.7]{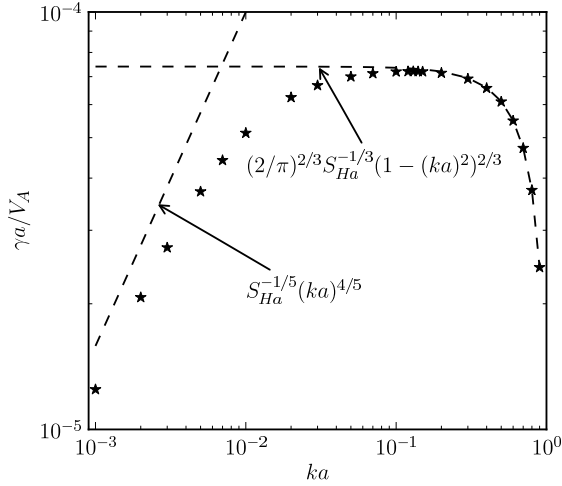}
\par\end{centering}

\caption{The dimensionless growth rate $\gamma a/V_{A}$ as a function of $ka$,
for a Harris sheet profile with $S_{Ha}=10^{12}$. Markers denote
values obtained by numerically solving the eigenvalue problem. Dashed
lines are theoretical predictions for the constant-$\psi$ and nonconstant-$\psi$
branches, Eqs. (\ref{eq:gamma-harris1}) and (\ref{eq:nonconst-psi-gamma}).
\label{fig:gamma-vs-ka}}

\end{figure}
\begin{figure}
\begin{centering}
\includegraphics[scale=0.7]{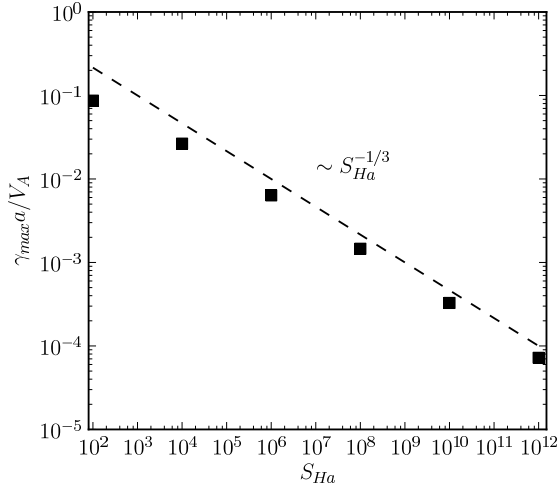}
\par\end{centering}

\caption{Scaling of the maximum dimensionless growth rate $\gamma_{max}a/V_{A}$
with $S_{Ha}$, for a Harris sheet profile. Markers denote values
obtained by numerically solving the eigenvalue problem. Dashed line
is the theoretical prediction $\gamma_{max}a/V_{A}\sim S_{Ha}^{-1/3}$.
\label{fig:gamma-vs-SHa}}
\end{figure}

These scalings are verified by numerically solving the eigenvalue
problem, Eqs. (\ref{eq:phi}) and (\ref{eq:psi}). Figure \ref{fig:gamma-vs-ka}
shows the dimensionless growth rate $\gamma a/V_{A}$ as a function
of $ka$, for a Harris sheet profile with $S_{Ha}=10^{12}$. The numerical
values agree with the analytic predictions remarkably well. Likewise,
Figure \ref{fig:gamma-vs-SHa} shows the scaling of the maximum dimensionless
growth rate $\gamma_{max}a/V_{A}$ with $S_{Ha}$. The numerical values
agree well with the theoretical scaling relation $\gamma_{max}a/V_{A}\sim S_{Ha}^{-1/3}$. 

Now we have all the ingredients for the linear theory of hyper-resistive
plasmoid instability. Substituting the hyper-resistive Sweet-Parker
current sheet width $\delta_{CS}\sim L/S_{H}^{1/4}$ for the current
sheet width $a$ in the linear tearing mode theory, we obtain the
following relation between $S_{H}$ and $S_{Ha}$ 
\begin{equation}
S_{Ha}\sim S_{H}^{1/4}.\label{eq:SH-SHa}
\end{equation}
Also the maximum growth rate scales as 
\begin{equation}
\gamma_{max}\sim S_{H}^{1/6}(V_{A}/L).\label{eq:gamma-max}
\end{equation}
As such, the instability growth rate increases for higher $S_{H}$,
similar to the resistive counterpart. The transition from constant-$\psi$
regime to nonconstant-$\psi$ regime occurs at $kL\sim S_{H}^{5/24}$.
Unlike the case of the resistive plasmoid instability, we are not
able to obtain a precise scaling for the number of plasmoids in the
hyper-resistive case, for the following reason. In the resistive plasmoid
instability, the scaling of the number of plasmoids can be inferred
from the wavenumber of the fastest growing mode, which coincides with
the transition wavenumber from the constant-$\psi$ regime to the
nonconstant-$\psi$ regime.\citep{LoureiroSC2007,BhattacharjeeHYR2009,HuangB2013}
That is not the case with hyper-resistivity. For hyper-resistive tearing
modes, the growth rate is approximately constant within the range
$S_{Ha}^{-1/6}\ll ka\ll1$ (see, for example, Figure \ref{fig:gamma-vs-ka}),
or equivalently within the range $S_{H}^{5/24}\ll kL\ll S_{H}^{1/4}$,
and the notion of fastest growing wavenumber loses its significance.

There are some subtleties in making qualitative comparisons between
the resistive and hyper-resistive plasmoid instabilities because the
mechanisms that break the frozen-in condition are quite different
in the two cases. A meaningful comparison may be made by rewriting
the scaling laws in terms of the aspect ratio $L/\delta_{CS}$ of
the primary Sweet-Parker current sheet, which is a common feature
for both models. For the resistive case, the maximum growth rate scales
as 
\begin{equation}
\gamma_{max}\sim\left(\frac{L}{\delta_{CS}}\right)^{1/2}\frac{V_{A}}{L},\label{eq:resistive1}
\end{equation}
and the fastest growing mode has a dimensionless wavenumber 
\begin{equation}
kL\sim\left(\frac{L}{\delta_{CS}}\right)^{3/4}.\label{eq:resistive2}
\end{equation}
On the other hand, for the hyper-resistive plasmoid instability, the
maximum growth rate scales as 
\begin{equation}
\gamma_{max}\sim\left(\frac{L}{\delta_{CS}}\right)^{2/3}\frac{V_{A}}{L},\label{eq:hyper1}
\end{equation}
and the growth rate peaks when $kL$ is within the range
\begin{equation}
\left(\frac{L}{\delta_{CS}}\right)^{5/6}\ll kL\ll\frac{L}{\delta_{CS}}.\label{eq:hyper2}
\end{equation}
From these scaling laws we may conclude that the hyper-resistive plasmoid
instability has a higher growth rate and prefers shorter wavelengths
as compared to the resistive one. Therefore, the hyper-resistive plasmoid
instability is even more explosive, and more efficient in generating
copious plasmoids, than the resistive plasmoid instability.

\section{Numerical Simulations\label{sec:Numerical-Simulations}}

\begin{figure}
\begin{centering}
\includegraphics[scale=0.8]{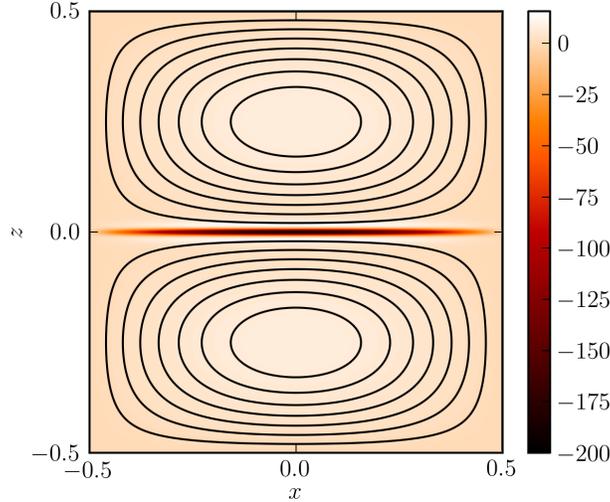}
\par\end{centering}

\caption{(Color online) The initial current density distribution, overlaid
with magnetic field lines.\label{fig:initial_condition}}

\end{figure}

\begin{figure}
\begin{centering}
\includegraphics[scale=0.8]{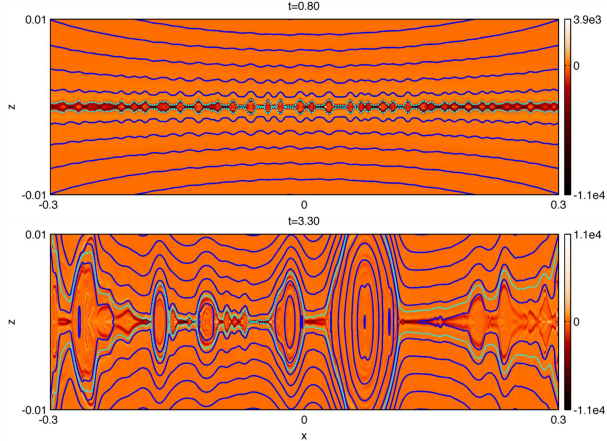}
\par\end{centering}

\caption{(Color online) Snapshots of the out-of-plane current profile, overlaid
with magnetic field lines, from a $S_{H}=10^{14}$ run. Dashed lines
represent separatrices separating the two primary merging islands
that drive the reconnection. The top panel shows the instability in
the early stage, and the bottom panel shows the fully developed nonlinear
stage. \label{fig:Snapshots}}

\end{figure}

The resistive plasmoid instability is of great interest because it
leads to a nonlinear regime where magnetic reconnection is drastically
different from the Sweet-Parker model. A pivotal question is how the
hyper-resistive plasmoid instability affects reconnection in the nonlinear
regime. To address this question, we employ the same simulation setup
of two coalescing magnetic islands as in Ref. \citep{HuangB2010}.
The governing equations are identical to the ones before, except that
resistivity is now replaced by hyper-resistivity. An isothermal equation
of state is assumed for simplicity. In normalized units, the simulation
box is a square in the domain $(x,z)\in[-1/2,1/2]\times[-1/2,1/2]$.
The initial magnetic field \textcolor{black}{is given by $\mathbf{B}_{0}=\nabla\psi_{0}\times\mathbf{\hat{y}}$,
where }$\psi_{0}=\tanh\left(z/h\right)\cos\left(\pi x\right)\sin\left(2\pi z\right)/2\pi$.
The parameter $h$, which is set to $0.01$ for all simulations, determines
the initial current layer width. The initial plasma density $\rho$
is approximately $1$, and the plasma temperature $T$ is $3$. The
density profile has a weak nonuniformity such that the initial condition
is approximately force-balanced. The initial peak magnetic field and
Alfv\'en speed are both approximately unity. Therefore, the hyper-resistive
Lundquist number $S_{H}=L^{3}V_{A}/\eta_{H}$ is simply $1/\eta_{H}$.
The plasma beta $\beta\equiv p/B^{2}=2\rho T/B^{2}$ is greater than
$6$ everywhere. Perfectly conducting and free slipping boundary conditions
are imposed along both $x$ and $z$ directions. Specifically, \textcolor{black}{we
have $\psi=0$, $\mathbf{u}\cdot\mathbf{\hat{n}}=0$, and $\mathbf{\hat{n}}\cdot\nabla\left(\mathbf{\hat{n}}\times\mathbf{u}\right)=0$
(here $\mathbf{\hat{n}}$ is the unit normal vector to the boundary).
Only the upper half of the domain ($z\ge0$) is simulated, and the
solutions in the lower half are inferred by symmetries. We employ
a uniform mesh along the $x$ direction, whereas the grid points along
the $z$ direction are strongly concentrated around $z=0$ to better
resolve the reconnection layer.} The highest resolution is 16000 grid
points along the $x$ direction and 1000 grid points along the $z$
direction, with the smallest grid size $\Delta z=1.8\times10^{-5}$.
Figure \ref{fig:initial_condition} shows the initial current density
distribution, overlaid with magnetic field lines. As the simulation
proceeds, the current layer first thins down and forms the primary
hyper-resistive Sweet-Parker layer. Subsequently, the primary current
layer may become unstable to the plasmoid instability if $S_{H}$
is above a threshold $S_{Hc}\simeq10^{10}$. We show snapshots of
a simulation with $S_{H}=10^{14}$ in Fig. \ref{fig:Snapshots}.

\subsection{Verification of Linear Theory}

\begin{figure}
\begin{centering}
\includegraphics[scale=0.7]{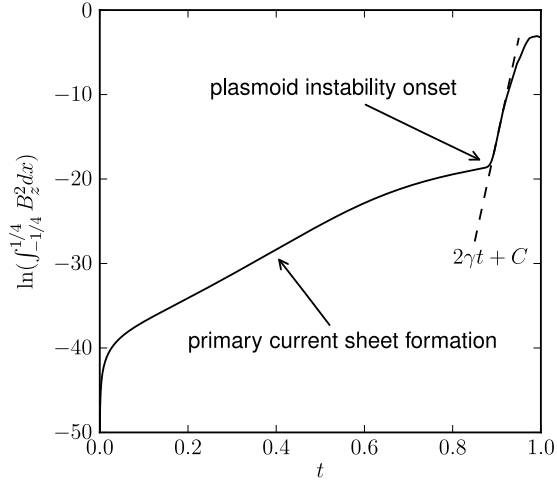}
\par\end{centering}

\caption{Measurement of the linear growth rate for the case $S_{H}=10^{14}$.\label{fig:growthrate-measurement}}
\end{figure}

\begin{figure}
\begin{centering}
\includegraphics[scale=0.7]{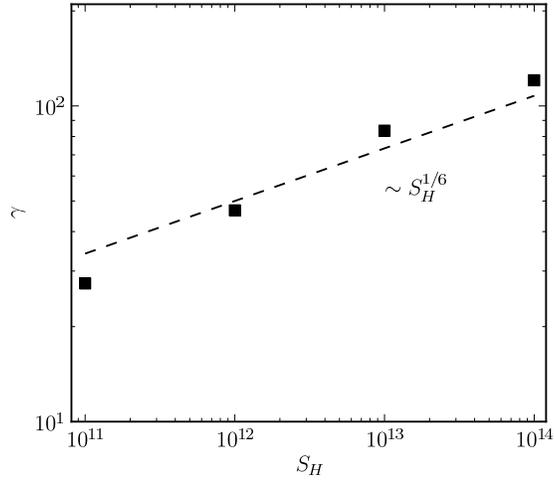}
\par\end{centering}

\caption{Scaling of the linear growth rate with respect to $S_{H}$. The dashed
line is the prediction from the linear theory. \label{fig:Scaing-of-growthrate}}

\end{figure}

Before we study magnetic reconnection in fully nonlinear regime for
this system, we first verify the prediction $\gamma_{max}\sim S_{H}^{1/6}(V_{A}/L)$
from the linear analysis. One difficulty in measuring the linear growth
rate is due to the fact that we do not start the simulation with a
Sweet-Parker current sheet. Rather, the Sweet-Parker current sheet
is established self-consistently during the current sheet thinning
phase. Therefore, the standard technique of adding a small perturbation
to an initial equilibrium and measuring the growth rate as the perturbation
grows does not apply here. The problem is that most physical quantities
evolve quite substantially during the current sheet thinning phase
before onset of the plasmoid instability, and it is difficult to filter
out the variations that are not due to the growth of the plasmoid
instability. This difficulty is overcome by looking at the component
$B_{z}$ along the central part of the reconnection layer $z=0$,
where \textbf{$B_{z}$} is identically zero initially and remains
small before onset of the plasmoid instability. After the onset, the
component $B_{z}$ develops fluctuations which rapidly grow as time
proceeds. To obtain the linear growth rate, we integrate $B_{z}^{2}$
at the central part of the current sheet along $z=0$, from $x=-1/4$
to $1/4$ at each time. The magnitude of $f(t)\equiv\int_{-1/4}^{1/4}B_{z}^{2}(t)dx$
remains small before onset of the plasmoid instability, and increases
abruptly after the onset. The linear growth rate $\gamma$ can be
obtained by fitting $\ln(f(t))$ to a linear function $\ln(f(t))\simeq2\gamma t+c$
during the period the plasmoid instability exhibits approximately
linear growth. This procedure is illustrated in Figure \ref{fig:growthrate-measurement}
for the case $S_{H}=10^{14}$. We measure the linear growth rates
for cases with $S_{H}$ ranging from $10^{11}$ to $10^{14}$. Figure
\ref{fig:Scaing-of-growthrate} shows the scaling of $\gamma$ with
respect to $S_{H}$. The results are in good agreement with the prediction
$\gamma_{max}\sim S_{H}^{1/6}(V_{A}/L)$.

\subsection{Scaling Laws in Nonlinear Regime}

\begin{figure}
\begin{centering}
\includegraphics[scale=0.7]{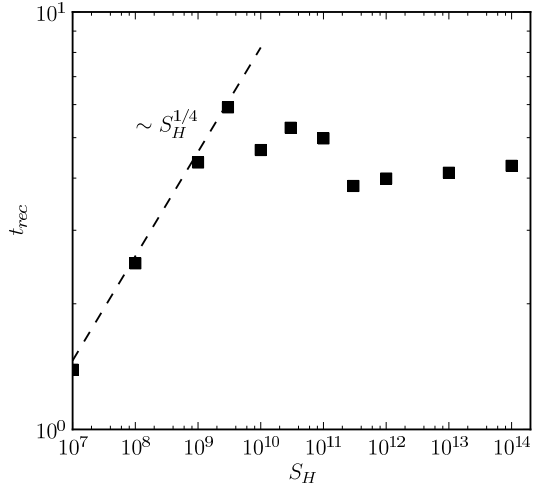}
\par\end{centering}

\caption{Scaling of the time to reconnect 25\% of the initial magnetic flux
with respect\label{fig:Scaling-of-reconnection-time} to $S_{H}$.}

\end{figure}

\begin{figure}
\begin{centering}
\includegraphics[scale=0.7]{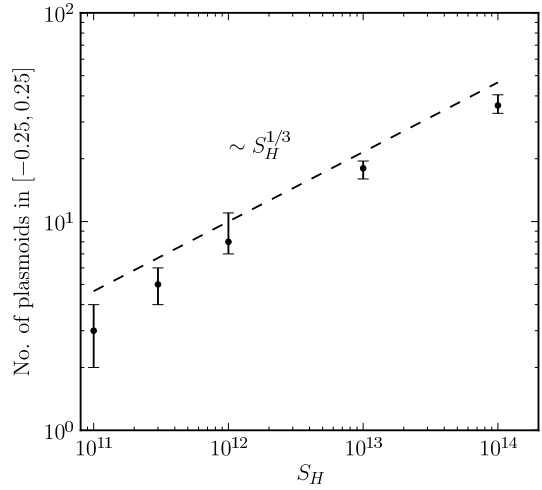}
\par\end{centering}

\caption{Scaling of the number of the plasmoids within $x\in[-0.25,0.25]$
with respect to $S_{H}$.\label{fig:Scaling-No-Plasmoids} }

\end{figure}

\begin{figure}
\begin{centering}
\includegraphics[scale=0.7]{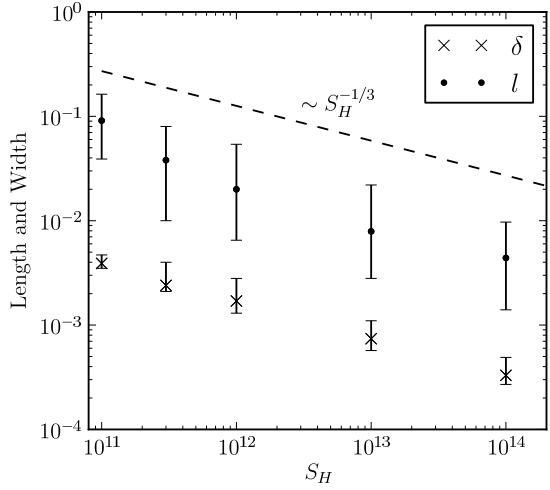}
\par\end{centering}

\caption{Scalings of the length $l$ and width $\delta$ of secondary current
sheets with respect to $S_{H}$.\label{fig:scaling-length} }

\end{figure}

\begin{figure}
\begin{centering}
\includegraphics[scale=0.7]{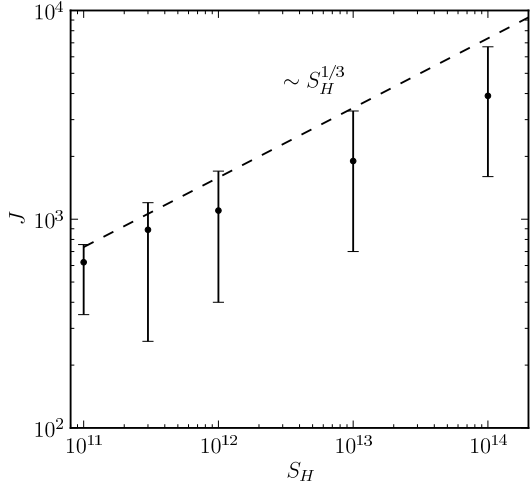}
\par\end{centering}

\caption{Scaling of the out-of-plane current density $J$ with respect to $S_{H}$.\label{fig:Jy-scaling}}
\end{figure}

\textcolor{black}{The next step is to establish scaling laws in the
nonlinear regime of hyper-resistive plasmoid instability, as we have
done in a previous study \citep{HuangB2010} for the resistive counterpart.
For nonlinear simulations, a low amplitude random forcing is included
to mimic thermal noise in real systems. In Ref. \citep{HuangB2010},
it is found that the result depends only weakly on the amplitude of
random forcing. For this reason, we set the random forcing amplitude
at a fixed level $\epsilon=10^{-4}$. The readers are referred to
Ref. \citep{HuangB2010} for details of how the amplitude $\epsilon$
is related to the energy input due the the random forcing and how
the random forcing is implemented numerically.}

\textcolor{black}{We employ the same diagnostics as in Ref. \citep{HuangB2010}.
To quantify the speed of reconnection, we measure the time it takes
to reconnect 25\% of the magnetic flux within the two merging islands,
which is denoted as $t_{rec}$. Figure \ref{fig:Scaling-of-reconnection-time}
shows the scaling of $t_{rec}$ with respect to $S_{H}$. For lower
$S_{H}$, the reconnection time scales as $t_{rec}\sim S_{H}^{1/4}$,
as expected from the hyper-resistive Sweet-Parker theory. When $S_{H}$
is above a critical value $S_{Hc}\simeq10^{10}$, the plasmoid instability
sets in and the reconnection time $t_{rec}$ becomes nearly independent
of $S_{H}$. In normalized units, the global characteristic values
for $V_{A}$ and $B$ are approximately 1, and $25\%$ of the initial
magnetic flux inside each of the islands is $0.04$, therefore the
normalized average reconnection rate is given by 
\begin{equation}
\frac{1}{BV_{A}}\left\langle \frac{d\psi}{dt}\right\rangle =\frac{0.04}{t_{rec}}.\label{eq:average-rate}
\end{equation}
In the regime $S_{H}>10^{10}$, $t_{rec}\simeq4$ to $5$ from our
simulations and the normalized reconnection rate is in the range $0.008$
to $0.01$. As such, the normalized reconnection rates here are on
par with those in resistive MHD models. \citep{BhattacharjeeHYR2009,CassakSD2009,HuangB2010}}

In \textcolor{black}{Ref. \citep{HuangB2010}, scaling laws for the
number of plasmoids, current sheet lengths and widths, and current
density have been deduced from simulation data. It was shown that
those scaling laws may be understood by a heuristic argument that
considers the reconnection layer as a chain of plasmoids connected
by marginally stable current sheets. The same argument may be carried
over to the hyper-resistive plasmoid instability, as follows. For
given $\eta_{H}$ and $V_{A}$, the critical length of a marginally
stable current layer is $L_{c}\sim(S_{Hc}\eta_{H}/V_{A})^{1/3}\sim L(S_{Hc}/S_{H})^{1/3}$
. Therefore we expect the number of plasmoids in the nonlinear regime
$n_{p}$ to scale like $n_{p}\sim L/L_{c}\sim(S_{H}/S_{Hc})^{1/3}$.
Furthermore, the width of the marginally stable current sheet $\delta_{c}\sim L_{c}/S_{Hc}^{1/4}\sim LS_{Hc}^{1/12}/S_{H}^{1/3}$,
and the current density $J\sim B/\delta_{c}\sim(B/L)S_{Hc}^{-1/12}S_{H}^{1/3}$.
Finally, we may estimate the reconnection rate by $\eta_{H}J/\delta_{c}^{2}\sim\eta_{H}B/\delta_{c}^{3}\sim BV_{A}/S_{Hc}^{1/4}$,
which is independent of $S_{H}$. This prediction of reconnection
rate being independent of $S_{H}$ is consistent with our results,
shown in Figure \ref{fig:Scaling-of-reconnection-time}. Likewise,
the predictions that the number of plasmoids scales as $S_{H}^{1/3}$,
the current sheet width and length both scale as $S_{H}^{-1/3}$,
and the current density scales as $S_{H}^{1/3}$ are also borne out
by our simulation data, shown in Figures \ref{fig:Scaling-No-Plasmoids},
\ref{fig:scaling-length}, and \ref{fig:Jy-scaling}. These data are
collected from time slices during the period to reconnect $25\%$
of the initial flux, and only plasmoids and secondary current sheets
within the domain $x\in[-0.25,0.25]$ are considered. Because the
number of plasmoids at a given snapshot varies in time, in Figure
\ref{fig:Scaling-No-Plasmoids} the medians are plotted, and the error
bars denote the first and third quartiles. Likewise, current sheets
also vary in length, width, and current density from one to another.
The data points and error bars in Fig. \ref{fig:scaling-length},
and Fig. \ref{fig:Jy-scaling} also denote the medians and the quartiles.}

\subsection{Statistical Distribution of Plasmoids}

\begin{figure}
\begin{centering}
\includegraphics[scale=0.8]{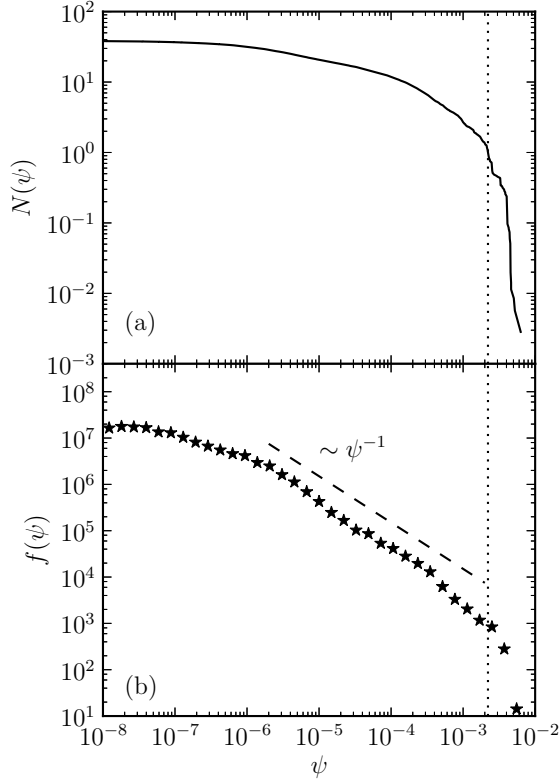}
\par\end{centering}

\caption{(a) Cumulative distribution function $N(\psi)$ and (b) probability
distribution function $f(\psi)$ from a $S_{H}=10^{14}$ simulation.
The vertical dotted line denotes where $N(\psi)=1$, indicating where
the dominant loss mechanism switches from coalescence to advection.\label{fig:pdf}}
\end{figure}

Seeking statistical descriptions of plasmoids has been a topic of
considerable interest in recent years, \citep{FermoDS2010,UzdenskyLS2010,FermoDSH2011,LoureiroSSU2012,HuangB2012,HuangB2013}
partly due to the possible link between plasmoids and energetic particles.\citep{DrakeSCS2006,ChenBPYBIMDLKVFG2008,DrakeSF2013}
In our recent work with resistive MHD, it was found numerically that
the distribution function $f(\psi)$ of magnetic flux $\psi$ inside
plasmoids exhibits a $f(\psi)\sim\psi^{-1}$ power-law distribution
over an extended range, followed by an exponential tail for large
plasmoids. A theoretical model was proposed that yields results consistent
with the numerical simulations.\citep{HuangB2012,HuangB2013} It has
been clarified that the transition from the power-law distribution
to the exponential tail is due to the dominant plasmoid loss mechanism
switching from coalescence to advection. This transition typically
occurs when the cumulative distribution function $N(\psi)\equiv\int_{\psi}^{\infty}f(\psi')d\psi'$
obeys the approximately inequality $N\lesssim1$, i.e. for the very
largest plasmoids in each snapshot. Because the theoretical model
only relies on the key assumption that secondary current sheets between
plasmoids are close to marginal stability and a few general assumptions
regarding coalescence and advection of plasmoids, the model can be
readily adapted to the case of hyper-resistive plasmoid instability.
Therefore, we expect plasmoids in hyper-resistive MHD model to follow
a similar distribution. That indeed appears to be the case. Figure
\ref{fig:pdf} shows the cumulative distribution function $N(\psi)$
and the distribution function $f(\psi)$ from a $S_{H}=10^{14}$ simulation.
The data set comprises 13486 plasmoids collected from 352 snapshots
during the period of reconnecting 25\% of the initial magnetic flux.
The distribution function $f(\psi)$ clearly exhibits an extended
$f(\psi)\sim\psi^{-1}$ power-law regime in the range between $\psi\sim10^{-6}$
and $\psi\sim10^{-3}$. Above $\psi\sim10^{-3}$ the distribution
makes a transition to a more rapid falloff. And this transition approximately
coincides the vertical dotted line, which denotes where $N(\psi)=1$,
indicating a switch of the dominant loss mechanism from coalescence
to advection. These features are qualitatively similar to the ones
with resistive plasmoid instability.

\section{Summary and Discussion\label{sec:Summary-and-Discussion}}

\begin{table}
\begin{centering}
\begin{tabular}{ccc}
\hline 
\hline  & Resistive  & Hyper-Resistive\tabularnewline
\hline 
$\gamma_{max}$ & $\sim\Lambda^{1/2}$ & $\sim\Lambda^{2/3}$\tabularnewline
$\kappa_{max}$ & $\sim\Lambda^{3/4}$ & $\Lambda^{5/6}\ll\kappa_{max}\ll\Lambda$\tabularnewline
$n_{p}$ & $\sim\Lambda^{2}$ & $\sim\Lambda^{4/3}$\tabularnewline
$\delta$ and $l$ & $\sim\Lambda^{-2}$ & $\sim\Lambda^{-4/3}$\tabularnewline
$J$ & $\sim\Lambda^{2}$ & $\sim\Lambda^{4/3}$\tabularnewline
Reconnection Rate & $\simeq10^{-2}V_{A}B$ & $\simeq10^{-2}V_{A}B$\tabularnewline
Plasmoid Distribution & $f(\psi)\sim\psi^{-1}$ & $f(\psi)\sim\psi^{-1}$\tabularnewline
\hline 
\end{tabular}
\par\end{centering}

\caption{Comparison between resistive and hyper-resistive plasmoid instabilities.
The scaling laws are expressed in terms of the aspect ratio $\Lambda=L/\delta_{CS}$.
Here $\gamma_{max}$ is the peak linear growth rate; $\kappa_{max}\equiv k_{max}L$
is the fastest growing wave number; $n_{p}$ is the number of plasmoids
in nonlinear regime; $\delta$ and $l$ are the thickness and length
of secondary current sheets; and $J$ is the current density. \label{tab:Comparison-of-scaling}}
\end{table}

In summary, we have carried out a linear instability analysis and
nonlinear simulations of the plasmoid instability when hyper-resistivity
is the mechanism of breaking field lines. We have found that the hyper-resistive
plasmoid instability is qualitatively similar to the resistive plasmoid
instability, although they follow different scaling laws both linearly
and nonlinearly. In the plasmoid-unstable regime, the reconnection
rate is found to be nearly independent of the hyper-resistive Lundquist
number $S_{H}$ instead of following the predicted $\sim S_{H}^{-1/4}$
scaling obtained by assuming a stable, extended current layer. The
reconnection rate in high-$S_{H}$ regime is approximately $0.01V_{A}B$,
which is similar to the value obtained with the resistive plasmoid
instability. The scaling laws in the nonlinear regime can be heuristically
derived by assuming secondary current sheets between plasmoids are
close to marginally stable, even though that assumption is clearly
oversimplified. The distribution of plasmoid magnetic flux is found
to obey a $f(\psi)\sim\psi^{-1}$ power law over an extended range,
followed by a rapid falloff for large plasmoids --- similar to the
result obtained for the resistive plasmoid instability.

Table \ref{tab:Comparison-of-scaling} summarizes the comparison between
resistive and hyper-resistive plasmoid instabilities. Here the scaling
laws are expressed in terms of the aspect ratio $\Lambda=L/\delta_{CS}$,
which is a common feature of both models. The aspect ratio scales
with respect to the resistive and hyper-resistive Lundquist numbers
as $\Lambda\sim S^{1/2}$ and $\Lambda\sim S_{H}^{1/4}$, respectively.
From these scaling relations, we can see that for the same aspect
ratio $\Lambda$, the hyper-resistive plasmoid instability has a higher
peak linear growth rate, and shorter wavelengths. Therefore, the hyper-resistive
plasmoid instability will set in more rapidly, with more plasmoids
at the early stage, compared to the resistive case. However, after
the plasmoid instability has developed into fully nonlinear regime,
more plasmoids will be present in the resistive case. The reason is
that resistivity is less effective in smoothing out small-scale structure,
which allows current sheet fragmentation to cascade down to deeper
level. 

The results in this paper may be relevant to the solar atmosphere,
where hyper-resistivity has been proposed as a possible mechanism
for corona heating.\citep{VanBallegooijenC2008} In addition, some
recent studies have found that current sheets formed after coronal
mass ejection (CME) events have thicknesses far broader than classical
or anomalous resistivity would predict, and it was suggested that
hyper-resistivity may be the cause.\citep{CiaravellaR2008,LinLKR2009}
Recently, a comparison of plasmoid distributions in post-CME current
sheets obtained from both solar observation and resistive MHD simulation
has been made.\citep{GuoBH2013} However, because plasmoid distributions
obtained from both resistive MHD and hyper-resistive MHD models are
essentially identical, statistical study of plasmoid distribution
alone will not be able to distinguish the two models. This conclusion
calls for other measures that can better tell apart different models.
The various scaling relations obtained in this paper may be able to
provide other insights on how this can be done.
\begin{acknowledgments}
\textcolor{black}{This work is supported by the Department of Energy,
Grant No. DE-FG02-07ER46372, under the auspice of the Center for Integrated
Computation and Analysis of Reconnection and Turbulence (CICART),
the National Science Foundation, Grant No. PHY-0215581 (PFC: Center
for Magnetic Self-Organization in Laboratory and Astrophysical Plasmas),
NASA Grant Nos. NNX09AJ86G and NNX10AC04G, and NSF Grant Nos. ATM-0802727,
ATM-090315 and AGS-0962698. YMH is partially supported by a NASA subcontract
to the Smithsonian Astrophysical Observatory's Center of Astrophysics,
Grant No. NNM07AA02C.  Computations were performed on facilities at
National Energy Research Scientific Computing Center. }
\end{acknowledgments}

\end{document}